\documentclass[12pt]{article}
\usepackage{bm,amsmath,amssymb,graphicx,mathrsfs}

\begin{document}
\pagenumbering{arabic}
\begin{titlepage}

\title{Probing the unitarity of the renormalizable theory of massive conformal 
gravity}

\author{F. F. Faria$\,^{*}$ \\
Centro de Ci\^encias da Natureza, \\
Universidade Estadual do Piau\'i, \\ 
64002-150 Teresina, PI, Brazil}

\date{}
\maketitle

\begin{abstract}
The presence of an unstable massive spin-$2$ ghost state in the renormalizable 
theory of massive conformal gravity leads to a pair of complex poles appearing 
in the first sheet of the energy plane. Here we show that the positions of these 
poles are gauge dependent, which makes the theory unitary.
\end{abstract}

\thispagestyle{empty}
\vfill
 \par
\bigskip
\noindent * felfrafar@hotmail.com \par
\end{titlepage}
\newpage

%%%%%%%%%%%%%%%%%%%%%%%%%%%%%%%%%%%%%%%%%%%%%%%%%%%%%%%%%%%%%%%%%%%%%%%%%%%%%%%

\section{Introduction}

%%%%%%%%%%%%%%%%%%%%%%%%%%%%%%%%%%%%%%%%%%%%%%%%%%%%%%%%%%%%%%%%%%%%%%%%%%%%%%%

The massive conformal gravity (MCG) is a renormalizable theory of gravity 
\cite{Faria1,Faria2} that has, in addition to the usual positive energy 
massless spin-$2$ field, a negative energy massive spin-$2$ field. 
At the quantum level, the negative energy field translates into an unstable 
negative norm ghost state. The instability of the ghost state makes it necessary 
to use a modified perturbation expansion in terms of dressed propagators. Since 
the bare propagator of the ghost state has a negative residue, the original 
real pole split into a pair of complex conjugate poles in the first riemannian 
energy sheet of the dressed propagator. If the positions of the complex poles 
are gauge-dependent, we can move them around by varying the corresponding gauge 
fixing parameter. In this case, the $S$-matrix connects only asymptotic 
states with positive norm and thus it is a unitary matrix.

It is well known that fourth order derivative theories of gravity have a 
massive ghost pole (or rather complex poles in the dressed propagator) whose 
position is gauge independent \cite{Johnston}. The advantage of MCG over 
these theories is that its linearized action is invariant, independently, 
under coordinate and conformal gauge transformations. Thus, even if the position 
of the MCG massive ghost pole is independent of the coordinate gauge fixing 
parameter, as happens in the fourth order derivative theories of gravity, its 
dependence on the conformal gauge fixing parameter alone is sufficient to 
ensure the unitarity of the theory.  

This paper is organized as follows. In Sect. 2, we describe the nature of 
the MCG massive ghost pole. In Sec. 3, we probe the gauge dependence of 
the positions of the MCG complex poles by using the Nielsen identities. In 
Sec. 4, we present our conclusions.

%%%%%%%%%%%%%%%%%%%%%%%%%%%%%%%%%%%%%%%%%%%%%%%%%%%%%%%%%%%%%%%%%%%%%%%%%%%%%%%

\section{Massive ghost pole}

%%%%%%%%%%%%%%%%%%%%%%%%%%%%%%%%%%%%%%%%%%%%%%%%%%%%%%%%%%%%%%%%%%%%%%%%%%%%%%%		

We start by considering the MCG action\footnote{Here we use units in which 
$c=\hbar=1$.} \cite{Faria3}
\begin{eqnarray}
S_{\textrm{MCG}} &=& \int{d^{4}x} \, \mathcal{L}_{\textrm{MCG}} 
\nonumber \\ &=& \frac{1}{k^2}\int{d^{4}x} \, \sqrt{-g}\left[ 
\varphi^{2}R + 6\partial_{\mu}\varphi\partial^{\mu}\varphi 
- \frac{1}{2m^2} C^{\alpha\beta\mu\nu}C_{\alpha\beta\mu\nu} \right],
\label{1}
\end{eqnarray}
where $k^2 = 16\pi G$, $m$ is a constant with dimension of mass, 
$\varphi$ is a scalar field called dilaton, 
\begin{equation}
C^{\alpha\beta\mu\nu}C_{\alpha\beta\mu\nu}  = R^{\alpha\beta\mu\nu}
R_{\alpha\beta\mu\nu} - 4R^{\mu\nu}R_{\mu\nu} + R^2 + 2\left(R^{\mu\nu}
R_{\mu\nu} - \frac{1}{3}R^{2}\right)
\label{2}
\end{equation}
is the Weyl tensor squared, $R^{\alpha}\,\!\!_{\mu\beta\nu}$ is the 
Riemann tensor, $R_{\mu\nu} = R^{\alpha}\,\!\!_{\mu\alpha\nu}$ is the 
Ricci tensor and $R = g^{\mu\nu}R_{\mu\nu}$ is the scalar curvature. 

Using the Lanczos identity, performing the background field expansions 
\begin{equation}
g_{\mu\nu} = \eta_{\mu\nu} + kh_{\mu\nu}, 
\label{3}
\end{equation}
\begin{equation} 
\varphi = 1 + k\sigma,
\label{4}
\end{equation}
and keeping only the terms of second order in the fields $h_{\mu\nu}$ 
and $\sigma$, it can be shown that the linear approximation of the MCG 
Lagrangian density (\ref{1}) is given by \cite{Faria1}
\begin{equation}
\bar{\mathcal{L}}_{\textrm{MCG}} = \bar{\mathcal{L}}_{EH} + 
2\sigma\bar{R} + 6 \partial^{\mu}\sigma\partial_{\mu}\sigma 
- \frac{1}{m^2}\left(\bar{R}^{\mu\nu}\bar{R}_{\mu\nu} 
- \frac{1}{3}\bar{R}^{2} \right),
\label{5}
\end{equation} 
where
\begin{equation}
\bar{R}_{\mu\nu} = \frac{1}{2} \left( \partial_{\mu}\partial^{\rho}
h_{\rho\nu} + \partial_{\nu}\partial^{\rho}h_{\rho\mu} 
- \Box h_{\mu\nu} 
- \partial_{\mu}\partial_{\nu}h  \right)
\label{6}
\end{equation}
is the linearized Ricci tensor,
\begin{equation}
\bar{R} =  \partial^{\mu}\partial^{\nu}h_{\mu\nu} - \Box h
\label{7}
\end{equation} 
is the linearized scalar curvature, and
\begin{equation}
\bar{\mathcal{L}}_{EH} = - \frac{1}{4} \Big( 
\partial^{\rho}h^{\mu\nu}\partial_{\rho}h_{\mu\nu} - 2\partial^{\mu}
h^{\nu\rho}\partial_{\rho}h_{\mu\nu} + 2\partial^{\mu}h_{\mu\nu}
\partial^{\nu}h - \partial^{\mu}h\partial_{\mu}h\Big)
\label{8}
\end{equation}
is the linearized Einstein-Hilbert Lagrangian density, with $\Box = 
\partial^{\mu}\partial_{\mu}$ and $h = \eta^{\mu\nu}h_{\mu\nu}$.

The linearized Lagrangian density (\ref{5}) is invariant under the coordinate 
gauge transformation
\begin{equation}
h_{\mu\nu} \rightarrow h_{\mu\nu} + \partial_{\mu}\chi_{\nu} + 
\partial_{\nu}\chi_{\mu},
\label{9}
\end{equation}
where $\chi^{\mu}$ is an arbitrary spacetime dependent vector field, and under 
the conformal gauge transformations
\begin{equation}
h_{\mu\nu} \rightarrow h_{\mu\nu} + 2\eta_{\mu\nu}\Lambda, \ \ \ \ \
\sigma \rightarrow \sigma - \Lambda,
\label{10}
\end{equation}
where $\Lambda$ is an arbitrary spacetime dependent scalar field. In order to 
fix these gauge freedoms, we must add the gauge fixing terms
\begin{equation}
\mathcal{L}_{GF1} = - \frac{1}{2\xi_{1}}\left( \partial^{\mu}h_{\mu\nu} 
- \frac{1}{2}\partial_{\nu}h \right)^{2},  
\label{11}
\end{equation}
\begin{equation}
\mathcal{L}_{GF2} =  \frac{1}{6\xi_{2}}\left( \frac{1}{m}\bar{R} - 
6m\xi_{2}\sigma\right)^{2},  
\label{12}
\end{equation}
to (\ref{5}), where $\xi_{1}$ and $\xi_{2}$ are gauge fixing parameters. 

Using the Barnes-Rivers projectors \cite{Barnes,Rivers}, and performing a 
long but straightforward calculation, we can show that the gauge fixed 
Lagrangian density $\bar{\mathcal{L}}_{\textrm{MCG}} + \mathcal{L}_{GF1} 
+ \mathcal{L}_{GF2}$ leads to the spin-$2$ part of the graviton bare propagator
\begin{equation}
D^{(2)}_{\mu\nu,\alpha\beta} = -i\left[ \frac{1}{p^2} -
\frac{1}{p^2 + m^{2}}\right]P^{(2)}_{\mu\nu,\alpha\beta}
\label{13}
\end{equation}
where
\begin{equation}
P^{(2)}_{\mu\nu,\alpha\beta} = \frac{1}{2}\left(\theta_{\mu\alpha}
\theta_{\nu\beta} +\theta_{\mu\beta}\theta_{\nu\alpha}\right) 
- \frac{1}{3}\theta_{\mu\nu}\theta_{\alpha\beta}
\label{14}
\end{equation}
is the spin-$2$ projector, with $\theta_{\mu\nu} = \eta_{\mu\nu} 
- p_{\mu}p_{\nu}/p^2$. The first term in the brackets in (\ref{13}) is the 
usual massless graviton pole at $p^2 = 0$ with positive residue, and the 
second is a massive ghost pole at $p^2 = -m^2 $ with negative residue. The 
massive ghost is, however, unstable because its mass is above the normal 
threshold of the massless graviton production. Thus, since the ordinary 
perturbation theory breaks down near the mass of an unstable particle 
\cite{Veltman}, we must use a modified perturbation series in which the 
bare propagator $D(p^2)$ is replaced by the dressed propagator \cite{Antoniadis}
\begin{equation}
\overline{D}(p^2) = \left[ D^{-1}(p^{2}) - \Pi(p^2) \right]^{-1},
\label{15}
\end{equation}
where $\Pi(k^2)$ is the sum of all one-particle irreducible (1PI) self-energy 
parts.

By coupling $N$ fermion fields to the MCG action (\ref{1}), expanding in powers 
of $1/N$, and using the Cauchy's integral theorem, we can write the spin-$2$ 
part of the graviton dressed propagator in the spectral form \cite{Tomboulis}
\begin{eqnarray}
\overline{D}^{(2)}_{\mu\nu,\alpha\beta} &=& -i\left[ \frac{1}{p^2} 
+ \frac{\mathscr{R}}{p^2 - M^{2}} + \frac{\mathscr{R^*}}{p^2 - M^{*2}} 
+ \frac{1}{2\pi}\int_{C}\frac{\rho(a)}{p^2-a}\, da\right]
P^{(2)}_{\mu\nu,\alpha\beta} 
\nonumber \\ && + \, \textrm{gauge terms},
\label{16}
\end{eqnarray}
where $M$, $M^*$, $\mathscr{R}$, and $\mathscr{R}^*$ are the positions and 
residues of a complex-conjugate pole pair, respectively, $\rho(a)$ is a 
spectral function obtained by cutting all the self-energy graphs of the 
continuum states, and $C$ is an appropriate path in the complex plane. We can 
see from (\ref{16}) that the pole for the unstable massive ghost has split into 
a pair of complex conjugate poles in the physical riemannian energy sheet, which 
supposedly breaks the unitarity of the $S$-matrix. However, if the positions of 
the complex poles are gauge-dependent, the unitarity of the gauge invariant 
$S$-matrix is satisfied. We will address this issue in the next section.

%%%%%%%%%%%%%%%%%%%%%%%%%%%%%%%%%%%%%%%%%%%%%%%%%%%%%%%%%%%%%%%%%%%%%%%%%%%%%%%

\section{Gauge dependence}

%%%%%%%%%%%%%%%%%%%%%%%%%%%%%%%%%%%%%%%%%%%%%%%%%%%%%%%%%%%%%%%%%%%%%%%%%%%%%%%

It is well known that for the position of a massive pole $m_{\textrm{pole}}^2$ 
to be gauge-independent, it must satisfy the Nielsen identity \cite{Nielsen}
\begin{equation}
\xi\frac{\partial m_{\textrm{pole}}^2}{\partial\xi} + C(\widehat{\phi},\xi)
\frac{\partial m_{\textrm{pole}}^2}{\partial\widehat{\phi}} = 0,
\label{17}
\end{equation}
where $\xi$ is any gauge fixing parameter, the hat represents setting all 
the fields of the theory, denoted generically by $\phi$, to their vacuum 
expectation values, and $C(\widehat{\phi},\xi)$ can be determined order by 
order in the loop expansion of the theory. 

In order to derive the Nielsen identities for the MCG complex poles, we 
choose the Becchi-Rouet-Stora-Tyutin (BRST) method 
\cite{Becchi1,Becchi2,Becchi3,Tyutin}, which consists in the inclusion of 
ghost and anti-ghost fields arising from the gauge redundancies of the theory 
\cite{Faddeev}. The addition of the compensating Faddeev-Popov ghost terms
\begin{equation}
\mathcal{L}_{FP1} =  \tilde{c}^{\mu}\Box c_{\mu},
\label{18}
\end{equation}
\begin{equation}
\mathcal{L}_{FP2} = 2 \, \tilde{c}\left( \Box - \xi_{2}m^{2}\right)c,  
\label{19}
\end{equation}
to $\bar{\mathcal{L}}_{\textrm{MCG}} + \mathcal{L}_{GF1} + \mathcal{L}_{GF2}$ 
gives the effective Lagrangian density
\begin{eqnarray}
\mathcal{L}_{\textrm{eff}} &=& \bar{\mathcal{L}}_{EH} + 
2\sigma\bar{R} + 6 \partial^{\mu}\sigma\partial_{\mu}\sigma 
- \frac{1}{m^2}\left(\bar{R}^{\mu\nu}\bar{R}_{\mu\nu} 
- \frac{1}{3}\bar{R}^{2} \right) \nonumber \\ && - \frac{1}{2\xi_{1}}
\left( \partial^{\mu}h_{\mu\nu} - \frac{1}{2}\partial_{\nu}h \right)^{2}
+ \frac{1}{6\xi_{2}}\left( \frac{1}{m}\bar{R} - 6\xi_{2}m\sigma\right)^{2}
\nonumber \\ &&
+ \tilde{c}^{\mu}\Box c_{\mu} + 2\tilde{c}\left( \Box - \xi_{2}m^{2}\right)c,
\label{20}
\end{eqnarray}
where ($\tilde{c}^{\mu}$)$c^{\mu}$ is a vector (anti-ghost)ghost field, and 
($\tilde{c}$)$c$ is a scalar (anti-ghost)ghost field. 

We can linearize the gauge fixing terms of the Lagrangian density (\ref{20}) 
by writing it in the form 
\begin{eqnarray}
\mathcal{L}_{\textrm{eff}} &=& \bar{\mathcal{L}}_{EH} + 
2\sigma\bar{R} + 6 \partial^{\mu}\sigma\partial_{\mu}\sigma 
- \frac{1}{m^2}\left(\bar{R}^{\mu\nu}\bar{R}_{\mu\nu} 
- \frac{1}{3}\bar{R}^{2} \right) \nonumber \\ &&  
- B^{\mu}\left( \partial^{\nu}h_{\mu\nu} - \frac{1}{2}\partial_{\mu}h \right)
+ \frac{1}{3}B\left( \frac{1}{m}\bar{R} - 6\xi_{2}m\sigma\right)
\nonumber \\ && + \frac{1}{2}\xi_{1}B^{\mu}B_{\mu} - \frac{1}{6}\xi_{2}B^2
+ \tilde{c}^{\mu}\Box c_{\mu} + 2 \tilde{c}\left(\Box - \xi_{2}m^{2}\right)c,
\label{21}
\end{eqnarray}
where
\begin{equation}
B_{\mu} = \frac{1}{\xi_{1}}\left( \partial^{\nu}h_{\mu\nu} 
- \frac{1}{2}\partial_{\mu}h \right)
\label{22}
\end{equation} 
is a Nakanishi-Lautrup auxiliary vector field, and 
\begin{equation}
B = \frac{1}{\xi_{2}}\left( \frac{1}{m}\bar{R} 
- 6\xi_{2}m\sigma\right)
\label{23}
\end{equation}
is a Nakanishi-Lautrup auxiliary scalar field. 

The action corresponding to the Lagrangian density (\ref{21}) is invariant under 
the BRST transformations
\begin{eqnarray}
&& \delta_{1}h_{\mu\nu} = \partial_{\mu}c_{\nu} + \partial_{\nu}c_{\mu}, 
\ \ \ \delta_{1}c_{\mu} = 0, \ \ \ \delta_{1}\tilde{c}_{\mu} = B_{\mu}, \ \ \
\delta_{1}B_{\mu} = 0, \nonumber \\ && \delta_{1}\sigma = 0, \ \ \ \ \ 
\delta_{1}c = 0, \ \ \ \ \ \delta_{1}\tilde{c} = 0, \ \ \ \ \ \delta_{1}B = 0,
\label{24}
\end{eqnarray}
\begin{eqnarray}
&& \delta_{2}h_{\mu\nu} = 2mc\eta_{\mu\nu}, \ \ \ \delta_{2}c_{\mu} = 0, \ \ \ 
\delta_{2}\tilde{c}_{\mu} = 0, \ \ \ \delta_{2}B_{\mu} = 0, \nonumber \\ && 
\delta_{2}\sigma = -mc, \ \ \ \ \ \delta_{2}c = 0, \ \ \ \ \ 
\delta_{2}\tilde{c} = B, \ \ \ \ \ \delta_{2}B = 0.
\label{25}
\end{eqnarray}
By adding the compensating terms
\begin{equation}
\mathcal{L}_{\Omega_1} = \Omega_{1}\left(- \frac{1}{2} B^{\mu}\tilde{c}_{\mu}
\right) = \Omega_{1}P_{1},
\label{26}
\end{equation}
\begin{equation}
\mathcal{L}_{\Omega_2} =  \Omega_{2}\left(\frac{1}{6} B\tilde{c}\right) 
= \Omega_{2}P_{2},  
\label{27}
\end{equation}
to (\ref{21}), we extend the BRST invariance of the theory to include the 
extra BRST transformations
\begin{equation}
\delta_{1}\xi_{1} = \Omega_{1}, \ \ \ \ \ \delta_{1}\xi_{2} = 0, 
\label{28}
\end{equation}
\begin{equation}
\delta_{2}\xi_{1} = 0, \ \ \ \ \ \delta_{2}\xi_{2} = \Omega_{2}, 
\label{29}
\end{equation}
where $\Omega_{1}$ and $\Omega_{2}$ are anticommuting constant scalar sources. 

The extended BRST invariance implies in the Ward-Takahashi identities 
\cite{Ward,Takahashi}
\begin{equation}
\Omega_{1}\frac{\partial \Gamma}{\partial\xi_{1}} +
\int{d^{4}x \left(\frac{\delta \Gamma}{\delta K^{\mu\nu}}\frac{\delta \Gamma}
{\delta \widehat{h}_{\mu\nu}} + \frac{\delta \Gamma}
{\delta\,\widehat{\tilde{c}}_{\mu}}\widehat{B}_{\mu} \right)} = 0,
\label{30}
\end{equation}
\begin{equation}
\Omega_{2}\frac{\partial \Gamma}{\partial\xi_{2}} +
\int{d^{4}x\left( \frac{\delta \Gamma}{\delta L^{\mu\nu}}\frac{\delta \Gamma}
{\delta \widehat{h}_{\mu\nu}}+ \frac{\delta \Gamma}{\delta L}
\frac{\delta \Gamma}{\delta\widehat{\sigma}} 
+ \frac{\delta \Gamma}{\delta\,\widehat{\tilde{c}}}\widehat{B} \right)} = 0,
\label{31}
\end{equation}
where $K^{\mu\nu}$, $L^{\mu\nu}$, and $L$ are sources for the composite 
fields $\delta_{1}h_{\mu\nu}$, $\delta_{2}h_{\mu\nu}$, and $\delta_{2}\sigma$, 
respectively, and $\Gamma$ is the effective action. Differentiating (\ref{30}) 
and (\ref{31}) with respect to $\Omega_{1}$ and $\Omega_{2}$, respectively, 
and setting $ \Omega_{1} = \Omega_{2} = 0$, we obtain
\begin{equation}
\frac{\partial \Gamma}{\partial\xi_{1}} -
\int{d^{4}x\left(\frac{\delta \Gamma(P_{1})}{\delta K^{\mu\nu}}\frac{\delta 
\Gamma}{\delta \widehat{h}_{\mu\nu}} + \frac{\delta \Gamma(P_{1})}
{\delta\,\widehat{\tilde{c}}_{\mu}}\widehat{B}_{\mu} \right)} = 0,
\label{32}
\end{equation}
\begin{equation}
\frac{\partial \Gamma}{\partial\xi_{2}} -
\int{d^{4}x\left( \frac{\delta \Gamma(P_{2})}{\delta L^{\mu\nu}}\frac{\delta 
\Gamma}{\delta \widehat{h}_{\mu\nu}}+ \frac{\delta \Gamma(P_{2})}{\delta L}
\frac{\delta \Gamma}{\delta \widehat{\sigma}} 
+ \frac{\delta \Gamma(P_{2})}{\delta\,\widehat{\tilde{c}}}\widehat{B} \right)}
= 0.
\label{33}
\end{equation}
Finally, by differentiating (\ref{32}) and (\ref{33}) twice with respect to 
$\widehat{h}_{\mu\nu}$, and setting all sources and fields equal to zero, with 
the exception of $\widehat{\sigma}$ which is taken to be a constant background 
field, we find the Nielsen identities for the inverse propagator
\begin{equation}
\left(\frac{\partial }{\partial\xi_{1}} \right)\frac{\delta^{2} 
\Gamma}{\delta \widehat{h}_{\mu\nu}\delta\widehat{h}_{\alpha\beta}} = 0,
\label{34}
\end{equation}
\begin{equation}
\left(\frac{\partial }{\partial\xi_{2}} 
+ C(\widehat{\sigma},\xi_{2})\frac{\partial }
{\partial\widehat{\sigma}} \right)\frac{\delta^{2} \Gamma}
{\delta \widehat{h}_{\mu\nu}\delta\widehat{h}_{\alpha\beta}} 
= \frac{\partial V(\widehat{\sigma},\xi_{2})}{\partial \widehat{\sigma}} 
\int{d^{4}x\,\frac{\delta^{3}\Gamma(P_{2})}{\delta L \delta 
\widehat{h}_{\mu\nu} \delta \widehat{h}_{\alpha\beta}}},
\label{35}
\end{equation}
where 
\begin{equation}
C(\widehat{\sigma},\xi_{2}) = -\int{d^{4}x\,\frac{\delta \Gamma(P_{2})}
{\delta L}},
\label{36}
\end{equation}
and $V(\widehat{\sigma},\xi_{2})$ is the effective potential, which is defined 
by
\begin{equation}
\Gamma[\widehat{\sigma}] = \int d^{4}x \, V(\widehat{\sigma},\xi_{2}).
\label{37}
\end{equation}

To find the Nielsen identities for the position of the complex massive pole 
$M^{2}$, we multiply (\ref{34}) and (\ref{35}) by $P^{(2)}_{\mu\nu,\alpha\beta}$ 
and set the external momentum $p^{2} = M^{2}$, which gives
\begin{equation}
\frac{\partial M^{2}}{\partial\xi_{1}} = 0,
\label{38}
\end{equation}
\begin{equation}
\frac{\partial M^{2}}{\partial\xi_{2}} + C(\widehat{\sigma},\xi_{2})
\frac{\partial M^{2}}{\partial\widehat{\sigma}} 
= \frac{(-i)\mathscr{R}}{10}\frac{\partial V(\widehat{\sigma},\xi_{2})}
{\partial \widehat{\sigma}} 
\left(P^{(2)}_{\mu\nu,\alpha\beta}\int{d^{4}x
\frac{\delta^{3}\Gamma(P_{2})}{\delta L \delta \widehat{h}_{\mu\nu} 
\delta \widehat{h}_{\alpha\beta}}}\right)\Bigg|_{p^{2}=M^{2}},
\label{39}
\end{equation}
where we used the inverse of the propagator (\ref{16}). We can see from 
(\ref{38}) that $M^{2}$ is independent of $\xi_{1}$. On the other hand, since 
there is no reason why the term between parentheses on the right-hand side of 
(\ref{39}) should vanish, $M^{2}$ will be independent of $\xi_{2}$ only if
\begin{equation}
\frac{\partial V(\widehat{\sigma},\xi_{2})}{\partial \widehat{\sigma}} = 0.
\label{40}
\end{equation}
However, it follows from (\ref{20}) that the zero order effective potential 
is given by
\begin{equation}
V_{0}(\widehat{\sigma},\xi_{2}) = 6\xi_{2}m^{2}\widehat{\sigma}^2.
\label{41}
\end{equation}
It is not difficult to see that the only solution of $\partial V_{0}/\partial
\,\widehat{\sigma} = 0$ is the trivial $\widehat{\sigma} = 0$. 
Thus, unless some unexpected cancellations take place in higher orders, we have
\begin{equation}
\frac{\partial V(\widehat{\sigma},\xi_{2})}{\partial \widehat{\sigma}} \neq 0,
\label{42}
\end{equation}
which means that $M^{2}$ depends on $\xi_{2}$. Similarly, we can show that 
the same is valid for $M^{*2}$.  Since we can define a gauge invariant 
$S$-matrix, such gauge-dependent poles disappear from the spectrum and 
unitarity is satisfied.

%%%%%%%%%%%%%%%%%%%%%%%%%%%%%%%%%%%%%%%%%%%%%%%%%%%%%%%%%%%%%%%%%%%%%%%%%%%%%%%

\section{Final remarks}

%%%%%%%%%%%%%%%%%%%%%%%%%%%%%%%%%%%%%%%%%%%%%%%%%%%%%%%%%%%%%%%%%%%%%%%%%%%%%%%

Here we presented a study on the unitarity of MCG. First, we noted that due to
the presence of the unstable massive spin-$2$ ghost state in the linearized 
theory, we must use a dressed propagator perturbation expansion, which results 
in the emergence of a pair of complex poles in the first sheet of the energy 
plane. Then, by using the Nielsen identities, it was shown that the positions 
of the complex poles are found to be dependent on the conformal gauge fixing 
parameter. This is enough to ensure that the excitations represented by the 
complex poles do not contribute to the gauge-invariant absorptive part of the 
$S$-matrix, leading to the unitarity of the theory. Therefore, we conclude 
that MCG is a consistent, renormalizable and unitary theory of quantum gravity.

%%%%%%%%%%%%%%%%%%%%%%%%%%%%%%%%%%%%%%%%%%%%%%%%%%%%%%%%%%%%%%%%%%%%%%%%%%%%%%%

\end{document}